\documentclass[twocolumn,showpacs,prl,aps]{revtex4-1}
\usepackage{bm,color,amsmath,amssymb,mathrsfs,latexsym,graphicx,psfrag}

\newcommand{\bs}[1]{\boldsymbol{#1}}
\newcommand{\be}{\begin{equation}}
\newcommand{\ee}{\end{equation}}
\newcommand{\bea}{\begin{eqnarray}}
\newcommand{\eea}{\end{eqnarray}}

\renewcommand{\phi}{\varphi}
\renewcommand{\epsilon}{\varepsilon}
\renewcommand{\vec}[1]{{\bf #1}}

\begin{document}

\title{Sublattice Interference in the Kagome Hubbard Model}

\author{Maximilian Kiesel${}^1$} 
\author{Ronny Thomale${}^2$} 
\affiliation{${}^1$Institute for Theoretical
  Physics, University of W\"urzburg, Am Hubland, D
  97074 W\"urzburg} 
\affiliation{${}^2$Department of Physics, Stanford University, Stanford, California 94305, USA}

\date{\today}

\begin{abstract}
We study the electronic phases of the kagome Hubbard model (KHM) in the weak coupling limit around van Hove filling. Through an analytic renormalization group analysis, we find that there exists a sublattice interference mechanism where the kagome sublattice structure affects the character of the Fermi surface instabilities. 
It leads to major suppression of $T_c$ for $d+id$ superconductivity in the KHM and causes an anomalous increase of $T_c$ upon addition of longer-range Hubbard interactions. We conjecture that the suppression of conventional Fermi liquid instabilities makes the KHM a prototype candidate for hosting exotic electronic states of matter at intermediate coupling.
\end{abstract}

\maketitle

{\it Introduction.} 
Understanding the variations of the critical scale $T_c$ of unconventional, i.e. electronically mediated, superconductivity is a long-standing challenge in condensed matter. For the cuprates, $T_c$ is non-universal and has been found to depend on various quantities such as structural parameters~\cite{jorgensen-96lnp1}, number of layers~\cite{iyo-07jpsj094711}, Fermi surface topology~\cite{pavarini-01prl047003}, and orbital content of electrons at the Fermi level~\cite{sakakibara-12prb064501}. For the latter, the $d_{z^2}$ admixture to the dominant $d_{x^2-y^2}$ Fermi surface character has been suggested as a substantial influence on $T_c$, a motif which is even more visible in the iron pnictides. There, at least all Fe  $t_{2g}$ orbitals ($d_{xz}$, $d_{yz}$, and $d_{xy}$) host large portions of electronic states in the vicinity of the Fermi surface, which generically necessitates a multi-band description. As a consequence, universal trends of pnictide superconductivity in terms of order parameter anisotropy and $T_c$ sensitively depend on the structural features which determine this orbital composition~\cite{kuroki-09prb224511,thomale-11prl187003}.  

\begin{figure*}[t]
  \begin{minipage}[l]{0.99\linewidth}
    \includegraphics[width=\linewidth]{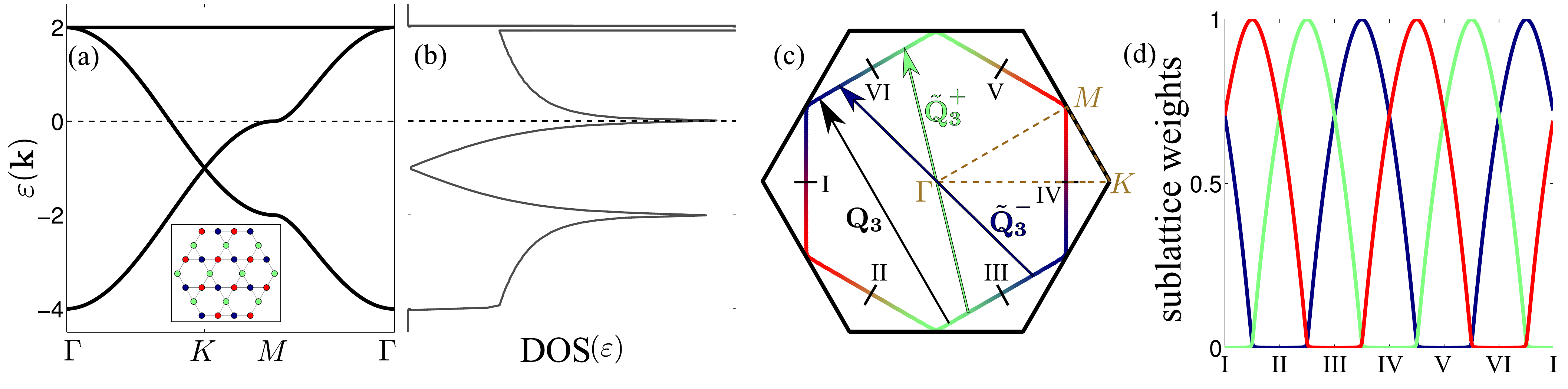}
  \end{minipage}
  \caption{(Color online). Fermi surface properties of the kagome tight binding model at $n=5/12$ total filling. (a) Band structure in units of $t$ resulting from the three sublattice structure of the kagome lattice (inset). The fillings $n=3/12$ and $n=5/12$  (dashed horizontal line) are located at van Hove singularities as visible in the density of state plot in (b). (c) The Fermi surface touches the $M$ point of the hexagonal Brioullin zone where the DOS is maximal; its topology allows for three nesting features one of which is $\vec{Q}_3=(-\pi/2,\sqrt{3}\pi/2)$. The colors blue, red, and green label the major sublattice occupation of the Fermi surface states. $\tilde{\vec{Q}}_3^{\pm}$ originate from opposite shifts of $\vec{Q}_3$ and link states of similar sublattice weights. (d) The FS labels I-VI defined in (c)  assist to read off the change of sublattice occupation weights $\vert u_s(\bs{k}) \vert$ along the Fermi surface.}
\label{fig1}
\vspace{-0pt}
\end{figure*}

Multi-band descriptions are both implied due to multiple orbitals and multiple sites associated with the unit cell of a given lattice. While previously mentioned superconductors are all square lattices with one single site per unit cell, the kagome lattice~\cite{elser89prl2405} possesses a minimal three-band model due to three sites per unit cell (inset Fig.~\ref{fig1}a). For the kagome Hubbard model (KMH), the three sublattices cause fundamental problems in characterizing its preferred electronic many-body phases. In the strong coupling limit at half filling, the kagome spin model exhibits strong quantum disorder fluctuations and has become one of the paradigmatic models of frustrated magnetism~\cite{ramirez94arms,misguichlhuillier,mendels-10jpsj011001}. While the associated Mott transition at finite coupling might still be described within dynamical mean field theory~\cite{ohashi-08ptps97}, the scope of collective electronic phases at intermediate Hubbard strength and general filling is particularly challenging to investigate: in the same way as electronic Bloch states at the Fermi level can involve different orbital admixtures for the mult-orbital case, the electronic states in the kagome lattice can be differently distributed among the multiple sublattices. 
From a tight binding perspective (Fig.~\ref{fig1}a), it is conceivable that the filling is a sensitive parameter in the KHM: we find two strongly dispersive bands and one flat band which, for appropriate fillings, has been suggested to be particularly susceptible to ferromagnetism along Stoner's criterion~\cite{tanaka-03prl067204}. 
While it is an ongoing demanding effort to identify kagome lattice materials at different electron fillings, a promising alternative route starts to emerge in optical kagome lattices of ultra-cold atomic gases, where the optical wavelengths can be suitably adjusted for fermionic isotopes such as ${}^6\text{Li}$ and ${}^{40}\text{K}$~\cite{jo-12prl045305}. 

In this paper, we take an itinerant viewpoint on the KHM and study how the tight-binding kagome model responds to weak local and longer range Hubbard interactions. The motivation is two-fold. First, we want to investigate what kind of competing Fermi surface instabilities emerge in the KHM revealing the interplay of the sublattice structure and Fermi surface topology. 
We particularly consider the regime of the dispersive bands around van Hove filling, where critical scales are enhanced due to large density of states and nesting becomes relevant (Fig.~\ref{fig1}). For superconductivity which we expect to find as the generically dominant instability channel for weak coupling~\cite{kohn-65prl524}, the multi-dimensional irreducible lattice representations associated with $C_{6v}$ symmetry on the kagome lattice which are even under inversion suggest the possibility of topological chiral singlet superconducting phases~\cite{kiesel}.
Second, the weak coupling limit equips us with a pivotal point of the KHM parameter space which we can solve up to analytic precision~\cite{raghu-10prb224505,raghu-12prb024516}. This provides a valuable starting point for subsequent effective studies at intermediate coupling, and as such improves our general understanding of the KHM.

%
%
%
%
%
%
%
%
%
%

\begin{figure}[t]
  \begin{minipage}[l]{0.99\linewidth}
    \includegraphics[width=\linewidth]{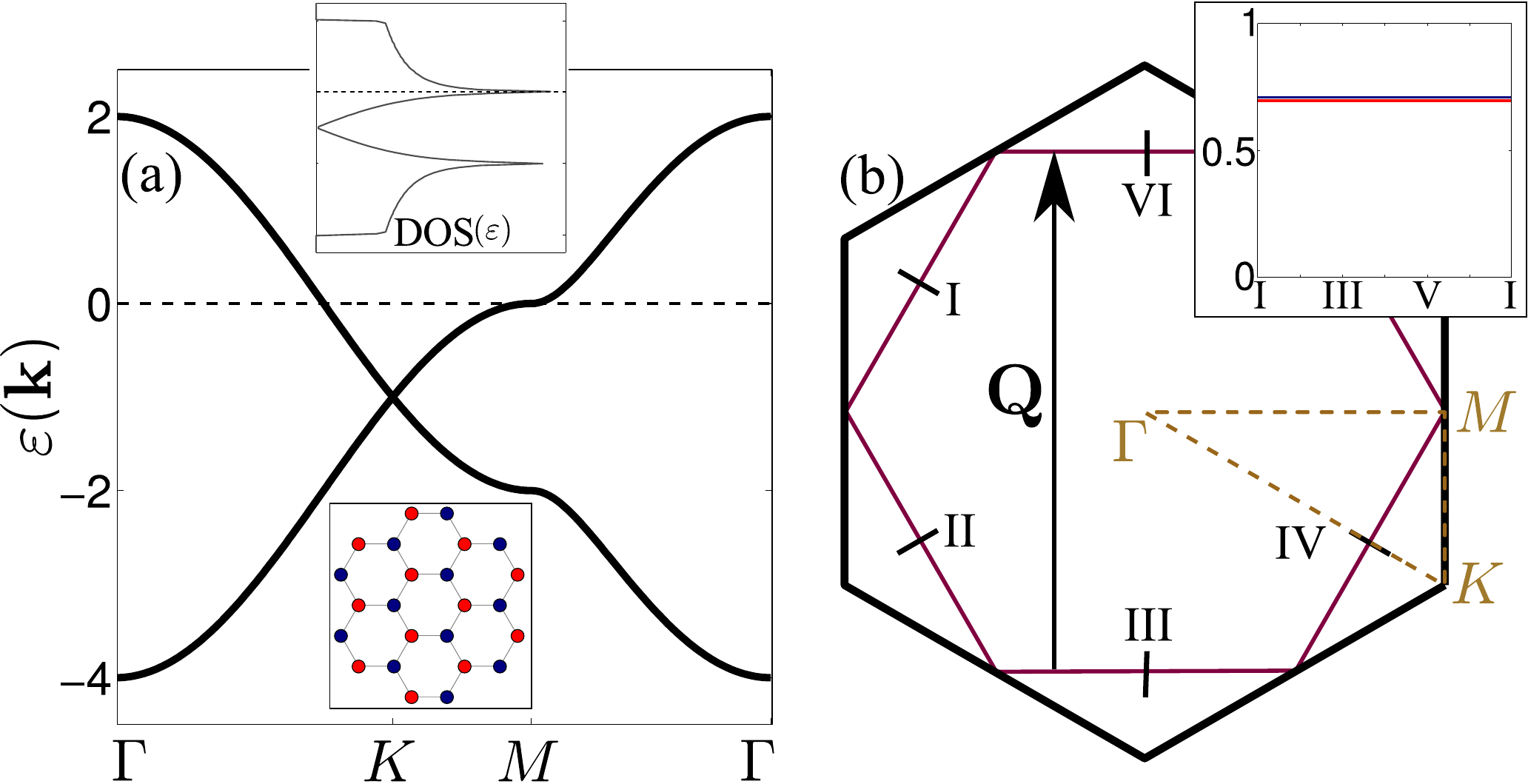}
  \end{minipage}
  \caption{(Color online). Fermi surface properties of the honeycomb tight binding model at $n=5/8$. (a) Band structure in units of $t$ along with the DOS (upper inset) and the two sublattice structure (lower inset).  Comparing the Fermi surface in (b) to the scenario in Fig.~\ref{fig1}, the Fermi surface topology and density state are approximately identical. 
Inset (b): the sublattice occupation along the Fermi surface is homogeneous.}
\label{fig2}
\vspace{-0pt}
\end{figure}

\begin{figure*}[t]
  \begin{minipage}[l]{0.99\linewidth}
    \includegraphics[width=\linewidth]{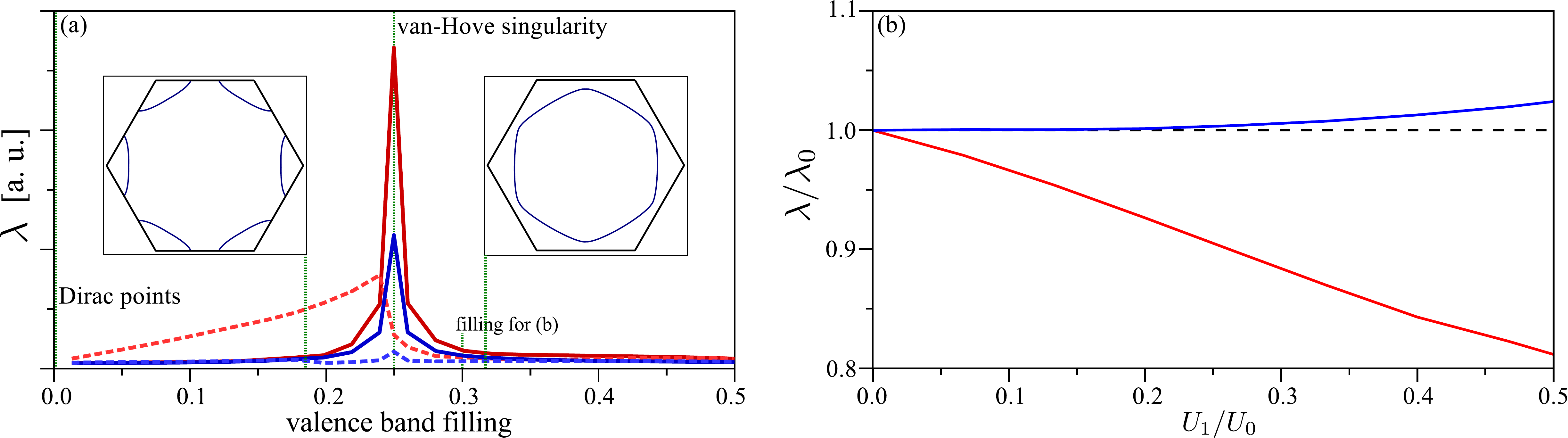}
  \end{minipage}
  \caption{(Color online). (a) Critical SC scale $\lambda$ versus valence band filling from weak coupling for local interaction only, presented for the kagome scenario in Fig.~\ref{fig1} (blue) and  the honeycomb scenario in Fig.~\ref{fig2} (red). Dashed line denotes $f$-wave, solid line $d+id$-wave. The valence band filling $n_v=0.25$ corresponds to a van Hove point. All scales in the kagome case are largely reduced as compared to the honeycomb case. Below $n_v$, the Fermi surface consists of disconnected pieces (left inset) and gives sizable $f$-wave for the honeycomb case. Above $n_v > 0.25$, $d+id$ is preferred along with a drop of $\lambda$ for larger $n_v$. (b) Relative change of $\lambda$ for finite $U_1$ as compared to the $U_0$ only case as a function of $U_1/U_0$ for both lattice scenarios at $n_v=0.3$.} 
\label{fig3}
\vspace{-0pt}
\end{figure*}

{\it Main results.} The sublattice structure of the kagome lattice has a crucial influence on the character of Fermi instabilities, as the sublattice distribution of electronic states varies along the Fermi surface (Fig.~\ref{fig1}).
We find $d+id$ superconductivity in the KHM in proximity at van Hove filling. This finding combined with the shape of the dispersive bands naively suggests a similarity to the honeycomb model doped to van Hove filling (Fig.~\ref{fig2}). There, $f$-wave is preferred at fillings where there are yet disconnected Fermi surface whereas $d+id$ is the leading instability as we find one Fermi pocket (Fig.~\ref{fig3}a). However, the scales of the KHM are suppressed as compared to the honeycomb scenario (Fig.~\ref{fig3}a): the KHM exhibits a mechanism which we call \textit{sublattice interference} affecting the emergence of Fermi surface instabilities, as the inhomogeneous sublattice distribution of Fermi level states causes reduced nesting effects. Furthermore, while the usual effect of long-range Hubbard interactions would be to reduce the critical scale of superconductivity~\cite{raghu-12prb024516,geballe}, it gets enhanced for the KHM as the long-range interactions help to relieve sublattice interference effects (Fig.~\ref{fig3}b).

{\it Model.} 
We consider the Hamiltonian
\begin{eqnarray}
&&H=H_0+H_{\text{int}}, \nonumber \\
&&H_0=t \sum_{\langle i, j \rangle} \sum_{\sigma} \left( c_{i\sigma}^\dagger c_{j\sigma}^{\phantom{\dagger}} + \text{h.c.} \right) + \mu\sum_{i,\sigma} n_{i,\sigma}, \label{h0} \\
&&H_{\text{int}}= U_0 \sum_{i}n_{i,\uparrow} n_{i,\downarrow} +\frac{U_1}{2}\sum_{\langle  i, j\rangle, \sigma, \sigma'} n_{i,\sigma} n_{j,\sigma'}, \label{hint}
\end{eqnarray}
where $c_{i,\sigma}$ denotes the annihilation operator of an electron at site $i$ with spin $\sigma$, $n_{i,\sigma}=c_{i\sigma}^\dagger c_{i\sigma}^{\phantom{\dagger}}$, and $\mu$ is the chemical potential which fixes the filling. While the local Hubbard interaction of scale $U_0$ is summed over all kagome sites, the sum of the tight binding model with energy scale $t$ as well as the nearest neighbor interaction of scale $U_1$ is summed over all neighbors on the kagome lattice which are on different sublattices (see inset in Fig.~\ref{fig1}a). In order to obtain the band structure of~\eqref{h0}, we perform the Fourier transform by dividing the kagome lattice into unit cells containing three sites each. This corresponds to quantum numbers of superlattice site $i$, sublattice $s$, and spin $\sigma$ characterizing the second quantized real space electron operators $c_{i,s,\sigma}^{(\dagger)}$. The diagonal form of~\eqref{h0} reads
\begin{equation}
H_0=\sum_{\bs{k},\sigma,n} \epsilon_n(\bs{k}) c_{\bs{k},n,\sigma}^{\dagger} c_{\bs{k},n,\sigma}^{\phantom{\dagger}},   \label{h0f}
\end{equation}
where $n$ denotes the band index and the band structure is shown in Fig.~\ref{fig1}a along with the density of states in Fig.~\ref{fig1}b. The transition from real space to momentum space upon Fourier transform reads
\begin{equation}
c_{i,s,\sigma}^{\dagger}=\sum_{\bs{k},n} u_{sn}^*(\bs{k}) c_{\bs{k},n,\sigma}^\dagger \exp(-i\bs{k}(\bs{R}_i+\bs{r}_s)), \label{uphase}
\end{equation} 
where $\bs{R}_i$ denotes the unit cell location and $\bs{r}_s$ the sublattice location within the unit cell. The core information which is relevant for investigating the interacting problem~\eqref{hint} is encoded in the transformation coefficients $u_{sn}(\bs{k})$ which we call sublattice weights in the following. For a given band $n$ and momentum point in the Brioullin zone $\bs{k}$, the coefficients obey $\sum_s \vert u_{sn} (\bs{k}) \vert^2=1$. 

{\it Method.} We employ perturbative renormalization group in the two-particle pairing channel to investigate the superconducting instabilities~\cite{raghu-10prb224505}. Summing over all diagrammatic contributions up to second order in the interaction scales $U_0$ and $U_1$, this approach is asymptotically exact for infinitesimal coupling for which the superconducting instabilities are found in the vicinity of the Fermi surface. The central quantity to compute for the perturbative RG is the pairing vertex
\begin{equation} 
\Gamma(\bs{k},\bs{p})\equiv \Gamma(\bs{k} \sigma,-\bs{k} \bar{\sigma},\bs{p} \sigma,-\bs{p} \bar{\sigma}),
\label{gamma}
\end{equation}
for incoming particles with momentum $\bs{k}$ and $-\bs{k}$ and outgoing particles with momentum $\bs{p}$ and $-\bs{p}$. Due to spin rotational invariance, the spin index effectively drops out of the pairing vertex as we can always constrain ourselves to the $S^z=0$ subblock where we obtain the triplet channel via antisymmetrization and the singlet channel via symmetrization of $\Gamma$. For infinitesimal interactions, all relevant momenta entering the interaction vertex are located at the Fermi surface. From the diagonal form of~\eqref{gamma}, we compute the superconducting instabilities and obtain the pairing vertex eigenvalues $\lambda_i$~\cite{raghu-10prb224505}, where $i=1, 2, \dots $ be ordered starting by $\lambda_1$ as the largest negative eigenvalue implying the strongest SC instability according to $T_c \sim E_{\text{F}} \exp [-1/\vert \lambda_1 \vert]$, were $E_{\text{F}}$ is the Fermi energy. Aside from the single-particle propagators as obtained by~\eqref{h0f}, the key object in the diagrammatic summation for $\Gamma(\bs{k},\bs{p})$ is the two-particle interaction vertex $V(\bs{k}_1\sigma,\bs{k}_2\bar{\sigma},\bs{k}_3\sigma,\bs{k}_4\bar{\sigma})$ with incoming particles $1$ and $2$ as well as outgoing particles $3$ and $4$, where $\bs{k}_4$ is given by momentum conservation, and, as for $\Gamma$, the spin index will effectively be dropped in the following.  

{\it Local Hubbard interaction.} Let us consider the case $U_1=0$ at van Hove filling $n=5/12$. The Fermi surface is depicted in Fig.~\ref{fig1}c. We only consider the interaction in the band at the Fermi level and hence drop the band index $n$ in the following. The interaction vertex takes the simple form
\begin{equation}
V(\bs{k}_1,\bs{k}_2, \bs{k}_3, \bs{k}_4)= U_0 \sum_s u_{s}^*(\bs{k}_1) u_{s}^*(\bs{k}_2) u_s(\bs{k}_3) u_s(\bs{k}_4). \label{vertex}
\end{equation}
From~\eqref{vertex}, because of the locality of $U_0$, the only momentum dependence is given by the sublattice distribution weights as defined in~\eqref{uphase}. Their evolution along the Fermi surface is depicted through color coding in Fig.~\ref{fig1}c and~\ref{fig1}d. Eq.~\ref{vertex} looks very familiar from orbital makeup factors in multi-orbital systems. In our case, this role is assigned to the sublattice weight distribution of the kagome model. As in the multi-orbital case, the sublattice now affects the nesting enhancement of particle-hole fluctuations along the Fermi surface. A first guess from Fermi surface topology without invoking the sublattice distribution would suggest the nesting vectors $\vec{Q}_1 = \pi \left( -\frac{1}{2} , -\frac{\sqrt{3}}{2} \right), \vec{Q}_2 = \pi \left( 1 , 0 \right), \vec{Q}_3 = \pi \left(-\frac{1}{2} , \frac{\sqrt{3}}{2} \right)$. As they connect Fermi surface points with mainly different sublattice occupation, however, the interaction vertex~\eqref{vertex} will be small as it is diagonal in the sublattice index $s$. This is what we call sublattice interference. In fact, because of sublattice interference, the most enhanced particle hole fluctuation channels split into $6$ different nesting vectors connecting equal sublattice weights. For $\vec{Q}_3$ this corresponds to a shift to $\tilde{\vec{Q}}_3^{\pm}=\vec{Q}_3\pm \pi \left( \frac{1}{4}, \frac{1}{4\sqrt{3}} \right)$ (Fig.~\ref{fig1}c). 

It is instructive to reconcile our findings with the Hubbard model on the honeycomb lattice with two lattice sites per unit cell (lower inset Fig.~\ref{fig2}), which has been recently investigated via RPA, 3-patch RG, weak coupling, and functional renormalization group~\cite{gonzalez08prb205431,honerkamp08prl146404,raghu-10prb224505,nandkishore-12np158,kiesel,wang-12prb035414}. There, the tight binding band structure matches with the dispersive bands of the kagome lattice and allows to similarly tune the honeycomb model to the equivalent van Hove filling (Fig.~\ref{fig2}a). While density of states (upper inset Fig.~\ref{fig2}a) as well as Fermi surface topology (Fig.~\ref{fig2}b) exactly match with the kagome case, the sublattice weights for the honeycomb model are homogeneous along the Fermi surface (inset Fig.~\ref{fig2}b), suggesting the absence of sublattice interference.

Fig.~\ref{fig3}a summarizes our results for local Hubbard interactions for the kagome and honeycomb tight binding model. We vary the doping around van Hove filling where we define the valence band filling $n_v$, i.e. the fraction of the partially occupied band, to enable a direct comparison of both cases. The van Hove filling is located at $n_v=1/4$, the Dirac cone filling at $n_v=0$. For $0<n_v<1/4$, the Fermi surfaces are disconnected (left inset Fig.~\ref{fig3}a), while they form one contingent pocket for $n_v>0.25$ (right inset Fig.~\ref{fig3}a). For the honeycomb scenario, we find that triplet $f$-wave SC is preferred for the former ($B_2$ representation of $C_{6v}$ symmetry group), while $d+id$-wave is preferred for the latter ($E_2$ representation). For the kagome scenario, we only find $d+id$-wave in close proximity to van Hove filling. In detail, for $d+id$ we find two degenerate SC eigenvalues $\lambda_{1,2}$ of $d$-wave symmetry, which then in any mean field treatment yield the preferential topological $d+id$ chiral superconducting state in order to avoid loss of condensation energy due to nodes which would necessarily cross with the Fermi surface~\cite{kiesel}. $d+id$ has also been obtained in variational cluster approximation calculations~\cite{yu-12prb144402} where, however, only local correlations are kept and no long-wavelength features of the electronic phases can be addressed.

Aside from the suppression of $f$-wave in the kagome case, the main difference between in the kagome ($k$) and honeycomb ($h$) scenario is seen in the quantitative difference of $\lambda$ (Fig.~\ref{fig3}a). At van Hove filling, $\lambda_{k}\sim 1/3 \lambda_{h}$. This illustrates at infinitesimal coupling how decisively  sublattice interference affects the emergence of superconductivity on the kagome lattice.

{\it Long-range Hubbard interactions.}
In the case of finite $U_1$, the interaction vertex gets significantly more complicated than for the onsite interaction scenario~\eqref{vertex}: momentum dependence now originates both from the harmonics associated with the finite interaction range as well as the sublattice weights. In particular, however, $V$ is not diagonal in the sublattice index anymore. We take a representative filling in the $d+id$ wave regime at $n_v=0.3$ and investigate the superconducting instabilities as a function of the ratio $U_1/U_0$ (Fig.~\ref{fig3}b). We plot the ratio $\lambda/\lambda_0$ where $\lambda_0$ is the pairing vertex eigenvalue at $U_1=0$. As elaborated on in~\cite{raghu-12prb024516}, the generic case which applies to the honeycomb scenario is such that long-range interaction should frustrate the pairing and induce a drop of $\lambda$, which might be tuned via the degree of external capacitive screening of the superconducting layer~\cite{geballe}. The KHM shows a notably different behavior, as $\lambda$ {\it increases} for longer range interactions. We can understand this phenomenon from the perspective of sublattice interference and the vertex function. As the vertex becomes non-diagonal in the sublattice index due to longer range interactions, this yields a reduction of sublattice interference effects as particle hole fluctuations between different sublattice component become sizable and allow for reestablishing nesting enhancement given by Fermi topology. Altogether, the reduction of sublattice interference effects overcompensates the effect of harmonic modulations due to the nearest neighbor term in~\eqref{hint} and yields a slight enhancement of $\lambda$ for long range Hubbard interactions (Fig.~\ref{fig3}b).

{\it Summary and outlook.} 
The KHM shows highly anomalous behavior in terms of weak coupling Fermi instabilities such as suppressed critical scales of superconductivity which increase upon addition of longer range Hubbard interactions. While this is beyond the scope of perturbative RG at infinitesimal coupling, our findings suggest that the KHM will likewise exhibit anomalously reduced critical scales of superconducting or spin density wave instabilities at intermediate coupling. This in turn might show the path towards stabilizing unconventional Fermi surface instabilities in the kagome Hubbard model.

\begin{acknowledgments}
  RT thanks S. A. Kivelson for various suggestions on the manuscript as well as S. Raghu for discussions on the RG approach in~\cite{raghu-10prb224505}. MK is supported by DFG-FOR 1162.  RT is supported by an SITP fellowship of Stanford University and by DFG-SPP 1458/1. 
\end{acknowledgments}


\end{document}